\documentclass[11pt,superscriptaddress,aps,prd,preprint]{revtex4}

\everymath{\displaystyle}
\usepackage{graphicx}

\usepackage[T1]{fontenc}
\usepackage{amsmath}
\usepackage{amssymb}
\usepackage{graphicx}
\usepackage{xcolor}

\newcommand{\bea}{\begin{eqnarray}}
\newcommand{\eea}{\end{eqnarray}}

\begin{document}


\title{Effects of the CPT-even and Lorentz violation on the Bhabha scattering at finite temperature}

\author{P. R. A. Souza}
\email{pablo@fisica.ufmt.br}
\affiliation{Instituto de F\'{\i}sica, Universidade Federal de Mato Grosso,\\
78060-900, Cuiab\'{a}, Mato Grosso, Brazil}

\author{A. F. Santos}
\email{alesandroferreira@fisica.ufmt.br}
\affiliation{Instituto de F\'{\i}sica, Universidade Federal de Mato Grosso,\\
78060-900, Cuiab\'{a}, Mato Grosso, Brazil}

\author{Faqir C. Khanna \footnote{Professor Emeritus - Physics Department, Theoretical Physics Institute, University of Alberta\\
Edmonton, Alberta, Canada}}
\email{khannaf@uvic.ca}
\affiliation{Department of Physics and Astronomy, University of
Victoria,BC V8P 5C2, Canada}

\begin{abstract}

In this paper a Lorentz-violating CPT-even non-minimal coupling term is considered. A new interaction term between fermions and photons emerges. In this context, the differential cross-section for Bhabha scattering at finite temperature is calculated. The temperature effects are introduced using the Thermo- Field Dynamics (TFD) formalism. It is shown that the differential cross-section is changed due to both effects, Lorentz violation and finite temperature.

\end{abstract}

\maketitle

\section{Introduction} 

General Relativity (GR) and  Standard Model (SM) of particle physics have been treated as separate parts of systems of particles for a long time. More recently it has been considered that it is probably a better idea to assume them as two components that belong to a unified theory. It is believed that a fundamental theory that combines SM and GR must emerge at very high energies, like the Planck scale, i.e. $\sim 10^{19}\,\mathrm{GeV}$. There are distinct areas seeking to unify quantum theory and GR within an underlying theory of quantum gravity, such as string theory \cite{st}, warped brane worlds \cite{bw}, loop quantum gravity \cite{lqg}, Horava-Lifshitz gravity \cite{hl}, $f(R)$ theory \cite{R1, R2, R3}, among others. Overall new physics has emerged. These theories lead to the possibility that small Lorentz and CPT symmetry violations may arise. An extension of the SM, known as Standard Model Extension (SME), has been constructed \cite{SME1, SME2}. The SME consist of two versions, the minimal and the non-minimal. The minimal SME contains only Lorentz-violating operators of mass dimensions three and four and in the non-minimal SME only operators of higher mass dimensions are considered. In general, the breakdown of SM is likely even though the effect is very small. The minimal contributions of SME with Lorentz and CPT violation will contribute to the results at low level.

Besides the studies carried out on the structure of SME, there are other ways to investigate the possibility of Lorentz violation. The Very Special Relativity (VSR) \cite{Cohen1, Cohen2} is a theory in which the laws of nature are not invariant under the whole Lorentz group but instead are invariant under subgroups of the Lorentz group that still preserves the basic elements of special relativity. Another way to introduce Lorentz violation is through non-minimal coupling terms. These terms modify the vertex interaction between fermions and photons. This new interaction term may be CPT-odd or CPT-even. Some applications using these non-minimal coupling terms have been developed \cite{Belich1, Belich2, SP, Casana1, Casana2, Casana3, Casana4, Our1, Our2}. In this paper, the CPT-even non-minimal coupling term is considered. Differential cross-section for Bhabha scattering at finite temperature is calculated. The Bhabha scattering consists of a scattering process between electron and positron with photon in the intermediate state. The calculation at finite temperature is carried using Thermo Field Dynamics (TFD).
  
The thermal effects may be introduced using the Matsubara formalism, a imaginary-time formalism \cite{Matsubara} or using the TFD formalism, a real-time formalism \cite{Umezawa1, Umezawa2, Kbook, Umezawa22, Khanna1, Khanna2}. Here the TFD formalism is considered. It is based on two fundamental elements: a) the doubling of Hilbert space and b) Bogoliubov transformation. The space is doubled to include usual Hilbet space and the tilde (dual) space. This leads to doubling of the degrees of freedom. The Green function is represented by a two dimensional structure.

The paper is organized as follows. In section II, the Lorentz violation theory is presented. In section III, the TFD formalism is introduced. The transition amplitude at finite temperature is obtained. In section IV, the differential cross-section for Bhabha scattering with Lorentz violating and CPT-even non-minimal coupling is calculated at finite temperature. In section V, some concluding remarks are presented. Throughout the paper, natural units, i.e. $c=\hbar=k_B=1$ are used.

\section{Lorentz-violating theory}

Here, a new CPT-even, dimension-five, nonminimal coupling connecting the fermionic and gauge fields is proposed in the Dirac equation, $(i\gamma^\mu {\cal D}_\mu-m)\psi=0$, where ${\cal D}_\mu$ is the nonminimal covariant derivative that leads to the Lorentz violation, which is defined as
\begin{equation}
{\cal D}_\mu=D_\mu+\frac{\lambda}{2}K_{\mu\nu\rho\alpha}\gamma^\nu F^{\rho\alpha}
\end{equation}
with $D_\mu=\partial_\mu+\imath eA_\mu$ being the usual covariant derivative, such that the standard relation $[D_\mu, D_\nu]=\imath e F_{\mu\nu}$ is satisfied, $e$ and $\lambda$ are the coupling constants and $K_{\mu\nu\rho\alpha}$ is a Lorentz-violating tensor that belongs to the CPT-even gauge sector of SME. It has the same symmetries as that of the Riemann tensor. In addition, this new coupling constant has mass dimension $[\lambda K_{\mu\nu\rho\alpha}]=-1$, which leads to a nonrenormalizable theory. However, this is not a problem for the investigation developed in this work since the interest is in the study of scattering processes at the tree-level. This tensor may be decomposed into birefringent and non-birefringent components. Here only the non-birefringent components are considered.   This Lorentz-violating tensor \cite{Kostelecky4} is 
\begin{equation}\label{3.2}
K_{\mu\nu\alpha\beta}=\frac{1}{2}\Big[g_{\mu\alpha}K_{\sigma\beta}-g_{\nu\alpha}K_{\mu\beta}+g_{\nu\beta}K_{\mu\alpha}-g_{\mu\beta}K_{\nu\alpha}\Big],
\end{equation}
where $K_{\mu\nu}$ is a symmetric and traceless rank-2 tensor.

Then the Lagrangian that describes the modified quantum electrodynamics is given as
\bea
{\cal L}&=& -\frac{1}{4}F_{\mu\nu}F^{\mu\nu}+\bar{\psi}\left(i\gamma^\mu {\cal D}_\mu-m\right)\psi.
\eea
The most important part of the Lagrangian for the study developed in this work is the interaction Lagrangian \cite{Casana, Our2}, which is given as
\begin{equation}\label{3.3}
\mathcal{L}_{_{\mathrm{SQED}}}^{I}=\mathcal{L}_{_{\mathrm{QED}}}^{I}+\lambda\overline{\Psi}\Big(\sigma_{\beta\nu} K^{\nu\mu}-\sigma^{\mu\nu} K_{\nu\beta}\Big)\kappa^\beta A_\mu\Psi
\end{equation}
with $\mathcal{L}_{_{\mathrm{QED}}}^{I}=-e\overline{\Psi}\gamma^\mu\Psi A_\mu$ being the usual QED interaction. It leads to the usual vertex, defined as $\Gamma^\mu_{(0)}=-\imath e\gamma^\mu$. The second term is a new interaction due to non-mininal coupling and the associated vertex is
\begin{equation}
\otimes\mapsto\mathrm{\Gamma}^\mu_{(1)}=-\imath\lambda\kappa^\beta\Big(\sigma_{\beta\nu} K^{\nu\mu}-\sigma^{\mu\nu} K_{\nu\beta}\Big)\label{3.4},
\end{equation}
with $\kappa^\beta$ being the 4-momentum of the photon. The new vertex components are
\begin{equation}\label{3.5}
\mathrm{\Gamma}^0_{(1)}=0,\quad\quad\quad \mathrm{\Gamma}^i_{(1)}=\mathrm{\Gamma}_{\!\!\mathrm{+is}}^\mathrm{i}+\mathrm{\Gamma}_{\!\!\mathrm{+an}}^\mathrm{i}+\mathrm{\Gamma}_{\!\mathrm{imp}}^\mathrm{i},
\end{equation}
where $\mathrm{\Gamma}_{\!\!\mathrm{+is}}^\mathrm{i}=-\imath\sqrt{s}\lambda K_{00}\sigma^\mathrm{0i}$ is the parity-even isotropic coefficient, $\mathrm{\Gamma}_{\!\!\mathrm{+an}}^\mathrm{i}=\imath\sqrt{s}\lambda K^\mathrm{ij}\sigma_\mathrm{0j}$ is the anisotropic parity-even part and $\mathrm{\Gamma}_{\!\mathrm{imp}}^\mathrm{i}=-\imath\sqrt{s}\lambda K_\mathrm{j}\sigma^\mathrm{ij}$ is the parity-odd component. Here $\kappa^\mu=(\sqrt{s},0)$ is used, where $\sqrt{s}$ is the energy in the center of mass. Since both coefficients provide contributions of the same order to the differential cross-section, here for simplicity only the parity-odd component is considered.

In this paper, the main objective is to calculate the differential cross-section for Bhabha scattering in the presence of a Lorentz-violating CPT-even non-minimal coupling at finite temperature. The thermal effects are introduced using the TFD formalism. The Feynman diagrams that describe this scattering process are shown in FIG. 1.
\begin{figure}[h]
 \centering
  \includegraphics[width=11cm]{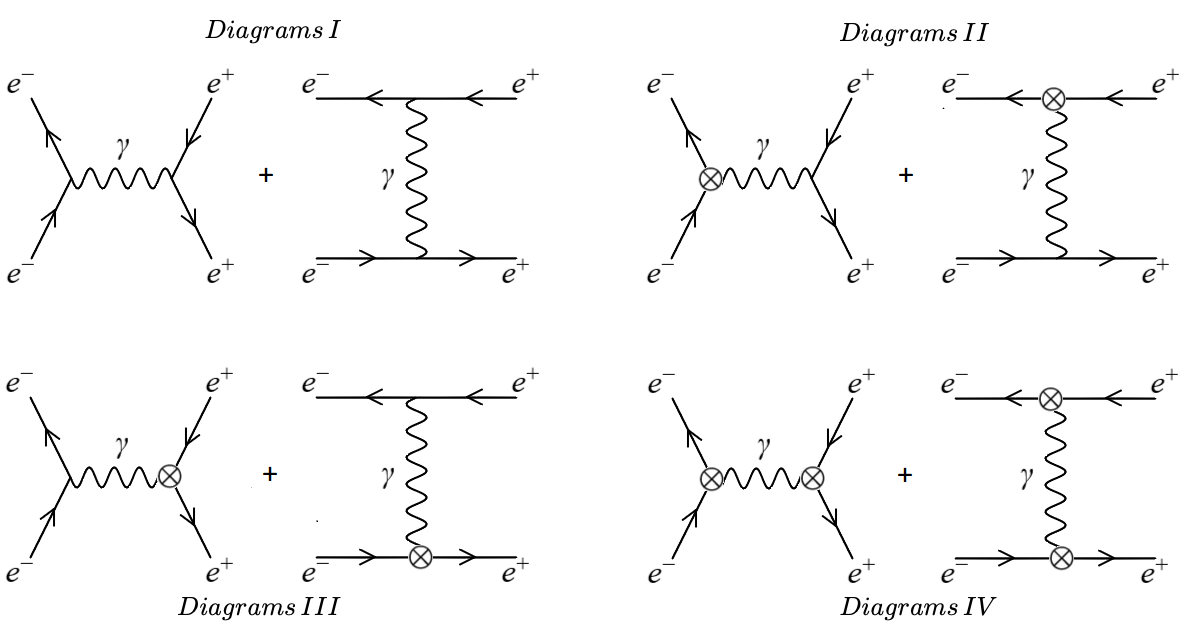}
   \caption{Feynman diagrams with different vertices. Each diagram is composed of the scattering process or t-channel (on the left side) and the annihilation process or s-channel (on the right side). Diagrams I represent the QED interaction, Diagrams II and III show the linear order for Lorentz-violating interaction and the Diagrams IV exhibit the quadratic order for Lorentz-violating interaction.}
    \label{v-bhabha}
\end{figure}

\section{TFD formalism and transition amplitude at finite temperature}

TFD is a real-time formalism of quantum field theory \cite{Umezawa1, Umezawa2, Kbook, Umezawa22, Khanna1, Khanna2}. In this formalism, the condition $\langle A \rangle=\langle0(\beta)|A|0(\beta)\rangle$ is satisfied, that is, the thermal average of any operator $A$ is equal to its temperature dependent vacuum expectation value. Here, $|0(\beta)\rangle$ a thermal vacuum state, with $\beta\propto T^{-1}$ and $T$ being the temperature. For this condition to be satisfied, two ingredients are needed: (1) the doubling
the degrees of freedom in the Hilbert space and (2) the Bogoliubov transformation. The Hilbert space is $S_T=S\otimes \tilde{S}$, with $S$ being the standard Hilbert space and $\tilde{S}$ the tilde (dual) space. The map between the tilde and non-tilde operators is defined by the tilde (or dual) conjugation rules. Considering two arbitrary operators $A$ and $B$, the conjugation rules are
\begin{equation}\label{2.1}
(A_i\tilde{)}=\tilde{A_i},\,\,\,(aA_i+b^*B_i\tilde{)}=(a^*\tilde{A}_i+b\tilde{B}_i),\,\,\,(A_i^{\dagger}\tilde{)}=\tilde{A}_i^{\dagger},\,\,\,
(A_iB_j\tilde{)}=\tilde{A}_i\tilde{B}_j,\,\,\,(\tilde{A}_i\tilde{)}=-\zeta A_i,
\end{equation}
with $\zeta = -1 (+1)$ for bosons (fermions). It is important to note that, the non-tilde operators act on physical states that belong to the usual Hilbert space and the tilde operators act in the dual Hilbert space.

The Bogoliubov transformation introduces the temperature effects. It consists in a rotation in the tilde and non-tilde variables. For fermions with $c_p^\dagger$ and $c_p$ being creation and annihilation operators respectively, Bogoliubov transformations are
\begin{eqnarray}
&& c_{p}={\mathtt u}(\beta)c_{p}(\beta)+{\mathtt v}(\beta)\tilde{c}^\dagger_{p}(\beta),\quad\quad c^\dagger_{p}={\mathtt u}(\beta)c^\dagger_{p}(\beta)+{\mathtt v}(\beta)\tilde{c}_{p}(\beta),\nonumber\\
&&\tilde{c}_{p}={\mathtt u}(\beta)\tilde{c}_{p}(\beta)-{\mathtt v}(\beta)c^\dagger_{p}(\beta),\quad\quad
\tilde{c}^\dagger_{p}={\mathtt u}(\beta)\tilde{c}^\dagger_{p}(\beta)-{\mathtt v}(\beta)c_{p}(\beta),\label{2.2}\
\end{eqnarray}
where $\mathtt{u}(\beta)$ and $\mathtt{v}(\beta)$ are given as
\begin{eqnarray}\label{2.3}
&&\mathtt{u}(\beta)=\cos(\theta(\beta))=(e^{-\beta|\kappa_0|}+1)^{-1},\\
&&\mathtt{v}(\beta)=\sin(\theta(\beta))=(e^{\beta|\kappa_0|}+1)^{-1},\nonumber\
\end{eqnarray}
with $\kappa_0$ being the energy. The anti-commutation relations for creation and annihilation operators are similar to those at zero temperature
\begin{equation}\label{2.4}
\{ c_{p}(\beta), c^\dagger_{q}(\beta) \}=\delta^3(p-q),\quad\quad\{ \tilde{c}_{p}(\beta), \tilde{c}^\dagger_{q}(\beta) \}=\delta^3(p-q),
\end{equation}
and other anti-commutation relations are null. For bosons, there are similar Bogoliubov transformations \cite{Kbook}.

The differential cross-section for Bhabha scattering at finite temperature is calculated using the transition amplitudes. In the TFD framework, the transition amplitude for an arbitrary scattering process is 
\begin{equation}\label{2.5}
\hat{\mathcal{M}}_{fi}(\beta)=\Big\langle f,\beta\Big|\hat{\mathcal{S}}\Big|i,\beta\Big\rangle,
\end{equation}
where the initial and final thermal states are
\begin{equation}\label{2.6}
|i,\beta\rangle=c^\dagger_{p_1}(\beta)d^\dagger_{p_2}(\beta)|0(\beta)\rangle, \quad\quad |f,\beta\rangle=c^\dagger_{p_3}(\beta)d^\dagger_{p_4}(\beta)|0(\beta)\rangle.
\end{equation}
The S-matrix $\hat{\mathcal{S}}$ in the duplicated space is defined as
\begin{equation}\label{2.7}
\hat{\mathcal{S}}=\sum_{n=0}^{\infty}\frac{(-\imath)^n}{n!}\int dx_1dx_2 \cdots dx_n\,\mathbb{T}\,[\hat{\mathcal{L}}_I(x_1)\hat{\mathcal{L}}_I(x_2)\cdots\hat{\mathcal{L}}_I(x_n)]\,,
\end{equation}
where $\mathbb{T}$ is the time ordering operator and $\hat{\mathcal{L}}_{I}(x)=\mathcal{L}_{I}(x)-\tilde{\mathcal{L}}_{I}(x)$ describes the interaction. Here up to the second order term of the S-matrix $\hat{\mathcal{S}}$ is considered. It is defined as
\begin{equation}\label{2.9}
\mathcal{S}^{(2)}=\frac{(-\imath)^2}{2}\int\!d^4xd^4y\,\mathbb{T}\,[\hat{\mathcal{L}}_I(x)\hat{\mathcal{L}}_I(y)]\,.
\end{equation}
Considering only the non-tilde part, since it describes the physical phenomenon, the transition amplitude at finite temperature becomes
\begin{equation}\label{2.10}
\mathcal{M}_{fi}(\beta)=\Big\langle f,\beta\Big|\mathcal{S}^{(2)}\Big|i,\beta\Big\rangle=\frac{(-\imath)^2}{2!}\!\!\int\!\!d^4xd^4y\Big\langle f,\beta\Big|\,\mathbb{T}\,[\mathcal{L}_I(x)\mathcal{L}_I(y)]\,\Big|i,\beta\Big\rangle.
\end{equation}

Using this transition amplitude, the differential cross-section at finite temperature for a scattering process is defined as
\begin{equation}\label{2.11}
\left(\frac{d\sigma}{d\Omega}\right)_{\!\!\beta}^2= \frac{\Big\langle\,\,\Big| \,\mathcal{M}_{fi}(\beta)\Big|^2\Big\rangle}{(8\pi E_{_{CM}})^2},
\end{equation}
where
\begin{equation}\label{2.12}
\Big\langle\,\,\Big| \mathcal{M}_{fi}(\beta)\Big|^2\Big\rangle=\frac{1}{4}\sum_{\mathrm{Spin}}\Big|\mathcal{M}_{fi}(\beta)\Big|^2.
\end{equation}
The average over the spin of incoming and outgoing particles is used.

In the next section, the differential cross-section for Bhabha scattering in the presence of a Lorentz-violating CPT-even non-minimal coupling at finite temperature is calculated.

\section{Differential Cross-Section at finite temperature}

The differential cross-section of the Bhabha scattering with Lorentz violation at finite temperature, using the center of mass (CM) frame is introduced. It is defined as
\begin{eqnarray}\label{4.1}
&&p_1=(E,p^i),\,\,\,p_2=(E,-p^{i}),\nonumber\\
&&p_3=(E,p'^i),\,\,\,\,p_4=(E,-p'^{i}),\\
&&\kappa=(p_1+p_2)=(\sqrt{s},0),\nonumber\
\end{eqnarray}
with $p_1$ and $p_2$ being the 4-momentum of incoming electrons and positrons while $p_3$ and $p_4$ are 4-momentum of outgoing electrons and positrons, respectively. Using the Mandelstam variables $s$, $t$ and $u$ that are defined as
\bea
s&\equiv&(p_1+p_2)^2=(p_3+p_4)^2,\nonumber\\ 
t&\equiv&(p_1-p_3)^2=(p_4-p_2)^2, \nonumber\\
u&\equiv&(p_1-p_4)^2=(p_3-p_2)^2,
\eea
we get
\begin{eqnarray}
(p_1\cdot p_2)&=&\frac{s}{2}-m^2,\quad\quad t=\Big(2m^2-\frac{s}{2}\Big)\Big(1-\cos(\theta)\Big)\nonumber\\
(p_1\cdot p_3)&=&m^2-\frac{t}{2},\quad\quad u=\Big(2m^2-\frac{s}{2}\Big)\Big(1+\cos(\theta)\Big)\nonumber\\
(p_1\cdot p_4)&=&m^2-\frac{u}{2},\quad\quad s+t+u = \sum_i m^2_i.
\end{eqnarray}

The fermion field is written as
\begin{equation}\label{4.3}
\Psi(\mathrm{x})=\int\!\!d\mathrm{p}\Big(c_\mathrm{p}u(\mathrm{p})e^{-\imath \mathrm{px}}+d_\mathrm{p}^\dagger v(\mathrm{p})e^{\imath \mathrm{px}}\Big),
\end{equation}
where $c_p$ and $d_p$ are annihilation operators for electrons and positrons, respectively and $u(\mathrm{p})$ and $v(\mathrm{p})$ are Dirac spinors. Then the transition amplitude becomes
\begin{equation}\label{4.4}
 \mathcal{M}_{\!fi\,}(\beta)=\frac{1}{2}\,\mathrm{\int\!d^4xd^4y}\sum_\mathrm{a,b}\Big\langle f,\beta\Big|\,\mathbb{T}\,\Big[\overline{\Psi}(\mathrm{x})\mathrm{\Gamma}_{(a)}^\mu\Psi(\mathrm{x})\overline{\Psi}(\mathrm{y})\mathrm{\Gamma}_{(b)}^\nu\Psi(\mathrm{y}) A_\mu(x)A_\nu(y)\,\Big]\,\Big|i, \beta\Big\rangle,
\end{equation}
with $a,b=0,1$ and  $\Gamma^\mu_{(a)}$ describes the vertices with $\Gamma^\mu_{(0)}$ representing the QED interaction and $\Gamma^\mu_{(1)}$ the new Lorentz-violating vertex.

Using Bogoliubov transformations, Eqs. (\ref{2.2}), the transition amplitude for $s$ and $t$ channels are
\begin{eqnarray}
&&\mathcal{M}^{(s)}_{(ab)}(\beta)=-\imath\sum_{spins}\frac{(\mathtt{u}^2(\beta)-\mathtt{v}^2(\beta))}{s}\Big[ \overline{u}(p_3)\Gamma_{(a)}^\mu v(p_4)\Big]\Big[ \overline{v}(p_2)\Gamma_{(b)\mu} u(p_1)\Big]\Delta(\kappa_2^2),\label{4.5}\\
&&\mathcal{M}^{(t)}_{(ab)}(\beta)=-\imath\sum_{spins}\frac{(\mathtt{u}^2(\beta)-\mathtt{v}^2(\beta))}{t}\Big[ \overline{u}(p_3)\Gamma_{(a)}^\mu u(p_1)\Big]\Big[ \overline{v}(p_2)\Gamma_{(b)\mu} v(p_4)\Big]\Delta(\kappa_1^2),\label{4.6}\
\end{eqnarray}
where $\vec{\kappa_1}$ and $\vec{\kappa_2}$ are the photon momentum of $t$ and $s$ channels, respectively and
\begin{eqnarray}
&& \Delta(\kappa_1)=1-\imath 2\pi t\frac{\delta(t)}{e^{\beta E_{_{CM}}}-1},\label{4.7}\\
&& \Delta(\kappa_2)=1-\imath 2\pi s\frac{\delta(s)}{e^{\beta E_{_{CM}}}-1}.\label{4.8}\
\end{eqnarray}
The photon propagator \cite{Umezawa22, Santos1} at finite temperature is defined as 
\begin{equation}\label{2.13}
\Big\langle 0(\beta)\Big|\mathbb{T}\,[A_\mu(\mathrm{x})A_\nu(\mathrm{y})] \Big|0(\beta)\Big\rangle=\imath\!\!\int\!\!\frac{d^4\kappa}{(2\pi)^4}
e^{-\imath\kappa\mathrm{(x-y)}}\frac{1}{\kappa^2}\Bigg[1-\imath\frac{2\pi\kappa^2\delta_{reg}(\kappa^2)}{e^{\beta|\kappa_0|}-1}\Bigg]\eta_{\mu\nu},
\end{equation}
with $\kappa$ being the photon propagator. It is important to note that, the photon propagator at finite temperature is a $2\times 2$ matrix. Here only the physical component which is given by the 11-component is considered.

The differential cross-section at finite temperature is given as
\begin{equation}\label{4.9}
\left(\frac{d\sigma}{d\Omega}\right)_{\!\!\beta}=\frac{\Big\langle\,\Big|\,\mathcal{M}_{fi}(\beta)\,\Big|^2\Big\rangle}{(8\pi E_{_{CM}})^2},
\end{equation}
with
\begin{equation}\label{4.10}
\Big\langle\,\Big|\,\mathcal{M}_{fi}(\beta)\,\Big|^2\Big\rangle=\Big\langle\,\Big|\,\mathcal{M}^{(s)}_{fi}(\beta)\Big|^2\Big\rangle + \Big\langle\,\Big|\,\mathcal{M}^{(t)}_{fi}(\beta)\Big|^2\Big\rangle - 2\mathbb{R}e\left\{\mathcal{M}^{(s)}_{fi}(\beta)\mathcal{M}^{\dagger(t)}_{fi}(\beta)\right\},
\end{equation}
where the last term corresponds to the cross term. Each term is calculated separately. Using the completeness relations:
\bea
\sum_{spin} u(p)\bar{u}(p)&=&\gamma^\mu p_\mu + m, \nonumber\\
\sum_{spin} v(p)\bar{v}(p)&=&\gamma^\mu p_\mu - m,
\eea
and the relation
\bea
\bar{v}(p_2)\gamma_\alpha u(p_1)\bar{u}(p_1)\gamma^\alpha v(p_2)=\mathrm{tr}\left[\gamma_\alpha u(p_1)\bar{u}(p_1)\gamma^\alpha  v(p_2)\bar{v}(p_2)\right],
\eea
the transition amplitude for $s$-channel is
\begin{eqnarray}
\Big\langle\,\,\Big|\,\mathcal{M}^{(s)}_{fi} \,\Big|^2\Big\rangle &=& \frac{1}{4}\sum_{a,b,c,d}\sum_{{Spins}}\Big |\mathcal{M}^{(s)}_{ ab}\mathcal{M}^{\dagger(s)}_{ cd} \Big |\nonumber\\
&=& \frac{\mathcal{S}(\beta)}{4s^2}\left\{ \sum_{a,b}\mathbb{S}^{\mu\nu}_{(ab)}\right\}\left\{\sum_{c,d}\mathbb{S}'_{(cd)\mu\nu}\right\},\label{4.15}\
\end{eqnarray}
where $a, b, c, d=0, 1$ are associated with the interactions terms and 
\bea
\mathcal{S}(\beta)&=&\tanh^4\left(\frac{\beta E_{_{CM}}}{2}\right)\left[ 1+ \frac{(2\pi)^2s^2\delta^2(s)}{(e^{\beta E_{_{CM}}}-1)^2}\right],\label{4.13}\\
\mathbb{S}^{\mu\nu}_{(ab)}&=&\mathrm{tr}\left\{ \Gamma_{(a)}^\mu\Big(p\!\!\!/ _4-m\Big)\overline{\Gamma}^\nu_{(b)} \Big(p\!\!\!/ _3+m\Big)\right\},\label{4.16}\\
\mathbb{S}'_{(cd)\mu\nu}&=&\mathrm{tr}\Big\{ \Gamma_{\mu (c)} \Big(p\!\!\!/ _1+m\Big)\overline{\Gamma}_{\nu(d)}\Big(p\!\!\!/ _2-m\Big)\Big\}.\label{4.17}
\eea
Then the differential cross-section at finite temperature associated with the $s$-channel is
\begin{equation}
\left(\frac{d\sigma^{(s)}}{d\Omega}\right)_{\!\!\beta}=\frac{\mathcal{S}(\beta)}{(8\pi E_{_{CM}})^2}\frac{1}{4s^2}\left\{ \sum_{a,b}\mathbb{S}^{\mu\nu}_{(ab)}\right\}\left\{\sum_{c,d}\mathbb{S}'_{(cd)\mu\nu}\right\}.\label{4.18}
\end{equation}

It is important to note that the indices $a, b$ assume two values, $0$ or $1$, corresponding to the usual and new non-minimal vertex. Using Eq (\ref{4.16}) and choosing $a=b=0$, we get
\begin{eqnarray}
&&\mathbb{S}^{\mu\nu}_{(00)}=4e^4\Big[ p_4^\mu p_3^\nu + p_4^\nu p_3^\mu - g^{\mu\nu}(p_3\cdot p_4)+g^{\mu\nu}m^2 \Big].\label{5s.1}\
\end{eqnarray}
This term contains only contributions due to the usual QED vertex. For $a=b=1$, the Lorentz-violating component is
\begin{eqnarray}
\mathbb{S}^{ij}_{(11)} &=& 4s\lambda^2\left\lbrace K^iK^j m^2-g^{ij}|\textbf{K}|^2m^2+p^i_4p^j_3|\textbf{K}|^2 + p^j_4p^i_3|\textbf{K}|^2 - p^i_4K^j(\textbf{K}\cdot\textbf{p}_3) +\right.\nonumber\\
&-& \left. p^j_4K^i(\textbf{K}\cdot\textbf{p}_3)- p^i_3K^j(\textbf{K}\cdot\textbf{p}_4)
- p^j_3K^i(\textbf{K}\cdot\textbf{p}_4) + 2g^{ij}(\textbf{K}\cdot\textbf{p}_3)(\textbf{K}\cdot\textbf{p}_4)\right.\nonumber\\
&+& \left. K^iK^j(\textbf{p}_3\cdot\textbf{p}_4) - g^{ij}|\textbf{K}|^2(\textbf{p}_3\cdot\textbf{p}_4)\right\rbrace. \label{5s.2}\
\end{eqnarray} 
The components $\mathbb{S}'_{(00)\mu\nu}$ and $\mathbb{S}'_{(11)ij}$ are obtained by the substitution $p_3\rightarrow p_1$, $p_4 \rightarrow p_2$ while the remaining components are zero, i.e., 
\begin{equation}\label{5s.3}
\mathbb{S}^{\mu\nu}_{(01)}=\mathbb{S}^{\mu\nu}_{(01)}=0,\,\,\,
\mathbb{S}^{0\nu}_{(11)}=\mathbb{S}^{\mu 0}_{(11)}=0.
\end{equation}
To the lowest order of the Lorentz-violating parameter, Eq. (\ref{4.18}) becomes
\begin{equation}\label{5s.4}
\left(\frac{d\sigma^{(s)}}{d\Omega}\right)_{\!\!\beta}\!\!\!\!=\frac{\mathcal{S}(\beta)}{(8\pi E_{_{CM}})^2}\,\frac{1}{4s^2}\,\left\{\mathbb{S}^{\mu\nu}_{(00)}\mathbb{S}'_{(00)\mu\nu} + 2\,\mathbb{S}^{ij}_{(00)}\mathbb{S}'_{(11)ij} \right\}.
\end{equation}
Using the CM frame, terms of the last equation are
\begin{eqnarray}
\mathbb{S}^{\mu\nu}_{(00)}\mathbb{S}'_{(00)\mu\nu} &=& 32e^4\left[
6m^4-m^2(s+t+u)+\frac{u^2+t^2}{4}
\right], \label{5s.5}\\
\mathbb{S}^{ij}_{(00)}\mathbb{S}'_{(11)ij} &=& -32e^2\lambda^2\left[
m^2\Big(2(\textbf{K}\cdot\textbf{p})^2 +m^2|\textbf{K}|^2\Big) +\right.\nonumber\\
 &-& \left. 2|\textbf{p}|^2\Big(2(\textbf{K}\cdot\textbf{p})^2-2|\textbf{K}|^2|\textbf{p}|^2\Big)
\right]. \label{5s.6}\
\end{eqnarray}
The differential cross-section corresponding to the $s$-channel is 
\begin{eqnarray}\label{5s.7}
\left(\frac{d\sigma^{(s)}}{d\Omega}\right)_{\!\!\beta} &=&
2\left(\frac{\alpha^2}{E^2_{_{CM}}}\right)\frac{\mathcal{S}(\beta)}{s^2}\left[\,\,\
\mathrm{I}_\lambda(m)-2\left(\frac{\lambda}{e}\right)^2|\textbf{p}|^2\Big(
2(\textbf{K}\cdot\textbf{p})^2-|\textbf{K}|^2|\textbf{p}|^2\Big)\right] +\nonumber\\
&+& \frac{1}{2}\left(\frac{\alpha^2}{E^2_{_{CM}}}\right)\mathcal{S}(\beta)\left\{\frac{u^2+t^2}{s^2}\right\},\
\end{eqnarray}
with $\alpha=e^2/4\pi$ and
\begin{equation}\label{5s.8}
\mathrm{I}_\lambda(m)=m^2
\left[
6m^2 -(s+t+u)
-\left(\frac{\lambda}{e}\right)^2
\Big(
2(\textbf{K}\cdot\textbf{p})^2+m^2|\textbf{K}|^2
\Big)
\right].
\end{equation}

Next the differential cross-section for the $t$-channel is given as
\bea
\left(\frac{d\sigma^{(t)}}{d\Omega}\right)_{\!\!\beta}&=&\frac{\mathcal{T}(\beta)}{(8\pi E_{_{CM}})^2}\,\frac{1}{4t^2}\left\{ \sum_{a,b}\mathbb{T}^{\mu\nu}_{(ab)}\right\}\left\{\sum_{c,d}\mathbb{T}'_{(cd)\mu\nu}\right\},\label{4.22}
\eea
with
\bea
\mathcal{T}(\beta)&=&\tanh^4 \left(\frac{\beta E_{_{CM}}}{2}\right)\left[ 1+ \frac{(2\pi)^2t^2\delta^2(t)}{(e^{\beta E_{_{CM}}}-1)^2}\right],\label{4.21}\\
\mathbb{T}^{\mu\nu}_{(ab)}&=&\mathrm{tr}\left\{ \Gamma_{(a)}^\mu\Big(p\!\!\!/ _3+m\Big)\overline{\Gamma}^\nu_{(b)}\Big(p\!\!\!/ _1+m\Big)\right\},\label{4.23}\\
\mathbb{T}'_{(cd)\mu\nu}&=&\mathrm{tr}\left\{ \Gamma_{\mu (c)} \Big(p\!\!\!/ _2-m\Big)\overline{\Gamma}_{\nu (d)} \Big(p\!\!\!/ _4-m\Big)\right\}.
\eea
Similar to the previous case, leads to
\begin{eqnarray}
\mathbb{T}^{\mu\nu}_{(00)}&=&4e^4\Big[ p_3^\mu p_1^\nu + p_1^\nu p_3^\mu - g^{\mu\nu}(p_1\cdot p_3)+g^{\mu\nu}m^2 \Big],\label{5t.1}\\
\mathbb{T}^{ij}_{(11)}&=& 4s\lambda^2\left\lbrace K^iK^j m^2-g^{ij}|\textbf{K}|^2m^2+p^i_3p^j_1|\textbf{K}|^2 + p^j_3p^i_1|\textbf{K}|^2 - p^i_3K^j(\textbf{K}\cdot\textbf{p}_1) +\right.\nonumber\\
&-& \left. p^j_3K^i(\textbf{K}\cdot\textbf{p}_1)- p^i_1K^j(\textbf{K}\cdot\textbf{p}_3)
- p^j_1K^i(\textbf{K}\cdot\textbf{p}_3) + 2g^{ji}(\textbf{K}\cdot\textbf{p}_1)(\textbf{K}\cdot\textbf{p}_3)\right.\nonumber\\
&+& \left. K^iK^j(\textbf{p}_1\cdot\textbf{p}_3) - g^{ij}|\textbf{K}|^2(\textbf{p}_1\cdot\textbf{p}_3)\right\rbrace, \label{5t.2}\
\end{eqnarray}
and
\begin{equation}\label{5t.3}
\mathbb{T}^{\mu\nu}_{(01)}=\mathbb{T}^{\mu\nu}_{(01)}=0,\,\,\,
\mathbb{T}^{0\nu}_{(11)}=\mathbb{T}^{\mu 0}_{(11)}=0.
\end{equation}
Then the differential cross-section in Eq. (\ref{4.22}) becomes
\begin{equation}\label{5t.4}
\left(\frac{d\sigma^{(t)}}{d\Omega}\right)_{\!\!\beta}\!\!\!\!=\frac{\mathcal{T}(\beta)}{(8\pi E_{_{CM}})^2}\,\frac{1}{4t^2}\left\lbrace \mathbb{T}^{\mu\nu}_{(00)}\mathbb{T}'_{(00)\mu\nu} + 2\,\mathbb{T}^{ij}_{(00)}\mathbb{T}'_{(11)ij} \right\rbrace,
\end{equation}
with
\begin{eqnarray}
\mathbb{T}^{\mu\nu}_{(00)}\mathbb{T}'_{(00)\mu\nu} &=& 32e^4\Big[
2m^4+m^2(t-u-s)+\frac{s^2+u^2}{4}
\Big], \label{5t.5}\\
\mathbb{T}^{ij}_{(00)}\mathbb{T}'_{(11)ij} &=& -32e^2\lambda^2\Big[
m^2\Big(2(\textbf{K}\cdot\textbf{p})^2 + m^2|\textbf{K}|^2\Big)+\nonumber\\
&-& 2|\textbf{p}|^2\Big(2(\textbf{K}\cdot\textbf{p})^2 + m^2|\textbf{K}|^2\Big)\Big].\label{5t.6}\
\end{eqnarray}
Using Eq. (\ref{5t.5}) and Eq. (\ref{5t.6}), the cross-section is 
\begin{eqnarray}
\left(\frac{d\sigma^{(t)}}{d\Omega}\right)_{\!\!\beta} &=&
2\left(\frac{\alpha^2}{E^2_{_{CM}}}\right)\frac{\mathcal{T}(\beta)}{t^2}\left[\,\,\
\mathrm{II}_\lambda(m)-2\left(\frac{\lambda}{e}\right)^2|\textbf{p}|^2\Big(
2(\textbf{K}\cdot\textbf{p})^2-|\textbf{K}|^2|\textbf{p}|^2\Big)\right] +\nonumber\\
&+& \frac{1}{2}\left(\frac{\alpha^2}{E^2_{_{CM}}}\right)\mathcal{T}(\beta)\left\{\frac{u^2+s^2}{t^2}\right\},\
\label{5t.7}
\end{eqnarray}
with
\begin{equation}\label{5t.8}
\mathrm{II}_\lambda(m)=m^2
\left[
2m^2-(s+u-t)-\left(\frac{\lambda}{e}\right)^2\Big(
2(\textbf{K}\cdot\textbf{p})^2+m^2|\textbf{K}|^2
\Big)
\right].
\end{equation}

Now the differential cross-section in the $u$-channel is 
\bea
\left(\frac{d\sigma^{(u)}}{d\Omega}\right)_{\!\!\beta}&=&-\frac{\mathbb{R}e\Big\{\mathcal{U}(\beta)\Big\}}{(8\pi E_{_{CM}})^2}\,\frac{1}{2ts}\!\!\sum_{a,b,c,d}\!\!\mathbb{U}_{\mathrm{(a\,b\,c\,d)}},
\eea
with
\bea
\mathbb{R}e\{\mathcal{U}(\beta)\}&=&\tanh^4\Big(\frac{\beta E_{_{CM}}}{2}\Big)\left[ 1 + \frac{(2\pi)^2s\delta(s)t\delta(t)}{(e^{\beta E_{_{CM}}}-1)^2} \right],\\
\mathbb{U}_{\mathrm{(abcd)}}&=&\mathrm{tr}\left\lbrace \Big(p\!\!\!/ _3+m\Big)
\Gamma^\mu_{(a)} \Big(p\!\!\!/ _4-m\Big)\overline{\Gamma}_{\nu\, (b)} \Big(p\!\!\!/ _2-m\Big)\Gamma_{\mu\, (c)}\Big(p\!\!\!/ _1+m\Big)\overline{\Gamma}^\nu_{(d)}\right\rbrace.
\eea

Considering the CM frame and lowest order of $\lambda$ parameter, it turns out that
\begin{eqnarray}
&&\mathbb{U}_{(0001)}=\mathbb{U}_{(0010)}=\mathbb{U}_{(0100)}=\nonumber\mathbb{U}_{(1000)}=0,\\
&& \mathbb{U}_{(1110)}=\mathbb{U}_{(1101)}=\mathbb{U}_{(0111)}=\mathbb{U}_{(1011)}=0,\label{5u.1}\\
&& \mathbb{U}_{(1001)}=\mathbb{U}_{(0011)}=\mathbb{U}_{(1100)}=0,
\end{eqnarray}
This leads to
\begin{equation}\label{5u.2}
\sum_{\mathrm{a,b,c,d}}\!\!\mathbb{U}_{\mathrm{(abcd)}}= \mathrm{U}_{(0000)}+\mathbb{U}_{(0110)}
+ \mathbb{U}_{(0101)}+\mathbb{U}_{(1010)},
\end{equation}
where the components are
\begin{eqnarray}
\mathbb{U}_{(0000)} &=& -32e^4\Big[
m^4 +\frac{m^2}{2}(s-t-3u)+\frac{u^2}{4}\Big],\label{5u.7}\\
\mathbb{U}_{(0110)} &=& -32e^2\lambda^2 s\, m^2\Big(
|\textbf{K}|^2|\textbf{p}|^2-(\textbf{K}\cdot\textbf{p})^2
\Big),\label{5u.8}\\
\mathbb{U}_{(0101)} &=& -32\frac{e^2\lambda^2 s}{2}\Big(2|\textbf{p}|^2-m^2\Big)\Big[
m^2-(\textbf{K}\cdot\textbf{p})^2
\Big],\label{5u.9}\\
\mathbb{U}_{(1010)} &=& -32\frac{e^2\lambda^2 s}{4}\Big[
|\textbf{p}|^2(\textbf{K}\cdot\textbf{p})^2+\label{5u.10}\\
&-& m^2\Big(
5(\textbf{K}\cdot\textbf{p})^2+m^2|\textbf{K}|^2 +
4|\textbf{K}|^2|\textbf{p}|^2
\Big)
\Big].\nonumber\
\end{eqnarray}
Then the differential cross-section for the interference term becomes
\begin{eqnarray}\label{5u.11}
\left(\frac{d\sigma^{(u)}}{d\Omega}\right)_{\!\!\beta}\!\!\!\! &=& 4\left(\frac{\alpha^2}{E^2_{_{CM}}}\right)\frac{\mathbb{R}e\Big\{\mathcal{U}(\beta)\Big\}}{ts}\Big\{\,
\mathrm{III}_\lambda(m) + \frac{s}{4}\left(\frac{\lambda}{e}\right)^2\times\\
&\times &\Big.\Big[\Big(|\textbf{p}|^2-m^2\Big)\Big(m^2 - (\textbf{K}\cdot\textbf{p})^2 \Big) + |\textbf{p}|^2(\textbf{K}\cdot\textbf{p})^2\Big]
\,\Big\}+\nonumber\\
&+& \frac{1}{2}\left(\frac{\alpha^2}{E^2_{_{CM}}}\right)\mathbb{R}e\Big\{\mathcal{U}(\beta)\Big\}\left\{2\frac{u^2}{ts}\right\}\nonumber\,
\end{eqnarray}
where
\begin{equation}\label{5u.12}
\mathrm{III}_\lambda(m) = m^2\left[
m^2+\frac{s-t-3u}{2} -\frac{s}{4}\left(\frac{\lambda}{e}\right)^2
\Big(
9(\textbf{K}\cdot\textbf{p})^2+m^2|\textbf{K}|^2
\Big)
\right].
\end{equation}

The total differential cross-section with Lorentz violation at finite temperature is
\begin{eqnarray}
\left(\frac{d\sigma}{d\Omega}\right)_{\!\!\beta}\!\!\!\!\! &=&\!\!\!\! \left(\frac{d\sigma^{(s)}}{d\Omega}\right)_{\!\!\beta}+\left(\frac{d\sigma^{(t)}}{d\Omega}\right)_{\!\!\beta}+\left(\frac{d\sigma^{(u)}}{d\Omega}\right)_{\!\!\beta}.\label{total}
\end{eqnarray}
Using Eqs. (\ref{5s.7}), (\ref{5t.7}) and (\ref{5u.11}), the last equation becomes
\begin{eqnarray}\label{5T.1}
\left(\frac{d\sigma}{d\Omega}\right)_{\!\!\beta}\!\!\!\!\!\! &=& \frac{1}{2}\left(\frac{\alpha^2}{E^2_{_{CM}}} \right)\left\{
\mathcal{S}(\beta)\left(\frac{u^2+t^2}{s^2}\right) + \mathcal{T}(\beta)\left(\frac{u^2+s^2}{t^2}\right) + \mathbb{R}e\Big\{\mathcal{U}(\beta)\Big\}\left(2\frac{u^2}{ts}\right) 
\right\}\nonumber\\
&& + 2\left(\frac{\alpha^2}{E^2_{_{CM}}}\right)
\left\{
\frac{\mathcal{S}(\beta)}{s^2}\left[
\mathrm{I}_\lambda(m)-2\left(\frac{\lambda}{e}\right)^2|\textbf{p}|^2\Big( 2(\textbf{K}\cdot\textbf{p})^2-|\textbf{K}|^2|\textbf{p}|^2\Big)
\right]+\right.\nonumber\\
&& + \frac{\mathcal{T}(\beta)}{t^2}\left[
\mathrm{II}_\lambda(m)-2\left(\frac{\lambda}{e}\right)^2|\textbf{p}|^2\Big( 2(\textbf{K}\cdot\textbf{p})^2-|\textbf{K}|^2|\textbf{p}|^2\Big)
\right]+ 2\frac{\mathbb{R}e\Big\{\mathcal{U}(\beta)\Big\}}{ts}\times\nonumber\\
&& \times\left.\left[
\mathrm{III}_\lambda(m) +
\frac{s}{4}\left(\frac{\lambda}{e}\right)^2\Big[\,
\Big(|\textbf{p}|^2-m^2\Big)\Big( m^2-(\textbf{K}\cdot\textbf{p})^2\Big)+|\textbf{p}|^2(\textbf{K}\cdot\textbf{p})^2
\,\Big]
\right]\right\}.
\end{eqnarray}
In this result product of delta functions with identical arguments are present. These are introduced due to the propagator at finite temperature \cite{Das1, Dolan, Das2, Das3}. A regularized  delta functions and their derivatives \cite{Van} are defined as
\bea
2\pi i\delta^n(x)=\left(-\frac{1}{x+i\epsilon}\right)^{n+1}-\left(-\frac{1}{x-i\epsilon}\right)^{n+1}.
\eea

It is important to note that Eq. (\ref{5T.1}) is an extended version of the usual QED with Lorentz violation at finite temperature. At very high temperatures, the thermal corrections are large and relevant. In addition, the cross-section of the Bhabha scattering is recovered within certain limits. For example, the limit $T\rightarrow 0 \,(\beta\rightarrow\infty)$ leads to $\mathcal{S}(0)=\mathcal{T}(0)=\mathbb{R}e\{\mathcal{U}(0)\}\rightarrow 1$ which gives the cross-section in the presence of the Lorentz-violating CPT-even non-minimal coupling at zero temperature. Furthermore, the  ultra-relativistic limit $(m\approx 0)$ and the case $\lambda\sim 0$ (Lorentz invariant case) leads to
\begin{equation}\label{5L.1}
\left(\frac{d\sigma}{d\Omega}\right)_{QED}= \left(\frac{\alpha^2}{2 E^2_{_{CM}}}\right)\left\{\frac{u^2+t^2}{s^2} + \frac{u^2+s^2}{t^2} + 2\frac{u^2}{ts} \right\}.
\end{equation}
This is the QED differential cross-section for Bhabha scattering.

\section{Conclusion}

This paper deals with Bhabha scattering using a non-minimal CPT-even Lorentz-violating coupling term. The covariant derivative is modified and a new interaction term appears. The Lagrangian consists of two parts: i) QED interaction and ii) the Lorentz violating CPT-even coupling term. In addition finite temperature is present in the system. The differential cross-section of Bhabha scattering is calculated. Furthermore thermal corrections to the cross-section are included using the real-time finite temperature effect. After completing this calculation, the Lorentz violation of the result is included so as to determine the overall violation effect. The cross-section is modified by both temperature as well as the Lorentz violation contribution. The results indicate the magnitude of the Lorentz violation compared to the role of the cross-section for a system without any violation effects on the system. There is no doubt that the Bhabha Scattering is very useful in providing such a behavior even though it is quite small indeed. It is important to note that the search for Lorentz violation using astrophysics has been carried out \cite{Astro}. Since the astrophysics processes occur at very high temperatures, the thermal effects may contribute to a new clash of constraints on Lorentz violation effects. In addition, this study shows that the presence of CPT-even Lorentz-violating parameter is almost as important as the case when the Lorentz violation remains intact.

\section*{Acknowledgments}

This work by A. F. S. is supported by CNPq projects 308611/2017-9 and 430194/2018-8.

\end{document}